\documentclass[prl,nofootinbib,twocolumn,superscriptaddress,preprintnumbers,
               balancelastpage,longbibliography,floatfix]{revtex4-1}

\usepackage{graphicx}
\usepackage{amssymb}
\usepackage{amsmath}
\usepackage{graphicx}
\usepackage{xcolor}
\usepackage{hyperref} 
\usepackage{bm}
\usepackage{chngcntr}

\usepackage[english]{babel}

\renewcommand{\vec}[1]{\mathbf{#1}}

\newcommand{\es}[2] {\begin{equation} \label{#1} \begin{split} #2 \end{split} \end{equation}}

\newcommand{\be}{\begin{equation}}
\newcommand{\ee}{\end{equation}}

\newcommand{\ev}[1]{\ensuremath{\langle #1 \rangle}}
\newcommand{\muas}{\ensuremath{\mu\text{as}}}

\begin{document}

\title{Stellar Wakes from Dark Matter Subhalos }

\preprint{MIT/CTP-4963, LCTP 17-04, MITP/17-079}

\author{Malte Buschmann}
\affiliation{Leinweber Center for Theoretical Physics, Department of Physics, University of Michigan, Ann Arbor, MI 48109 USA}
\affiliation{PRISMA Cluster of Excellence and Mainz Institute for Theoretical Physics, Johannes Gutenberg University, 55099 Mainz, Germany}

\author{Joachim Kopp}
\affiliation{PRISMA Cluster of Excellence and Mainz Institute for Theoretical Physics, Johannes Gutenberg University, 55099 Mainz, Germany}

\author{Benjamin R. Safdi}
\affiliation{Leinweber Center for Theoretical Physics, Department of Physics, University of Michigan, Ann Arbor, MI 48109 USA}

\author{Chih-Liang Wu}
\affiliation{Center for Theoretical Physics, Massachusetts Institute of Technology, Cambridge, MA 02139, U.S.A.}

\date{\today}

\begin{abstract}
We propose a novel method utilizing stellar kinematic data to detect low-mass substructure in
the Milky Way's dark matter halo. By probing characteristic wakes that
a passing dark matter subhalo leaves in the phase space distribution of ambient
halo stars, we estimate sensitivities down to subhalo masses $\sim
10^7\,M_\odot$ or below.  The detection of such subhalos would have
implications for dark-matter and cosmological models that predict modifications
to the halo-mass function at low halo masses.  We develop an analytic formalism
for describing the perturbed stellar phase-space distributions, and we demonstrate through simulations the ability to detect subhalos using the phase-space model and a likelihood framework.
Our method complements existing
methods for low-mass subhalo searches, such as searches for gaps in stellar
streams, in that we can localize the positions and velocities of the subhalos
today.   
\end{abstract}

%---------------------------------------------------------------------------
\maketitle
%---------------------------------------------------------------------------

\noindent 
{\bf Introduction.}---The Standard Cosmological Model ($\Lambda$CDM), based on
cold dark matter (CDM) and a cosmological constant ($\Lambda$), predicts that
the otherwise homogeneous primordial plasma features small density
perturbations with a nearly scale invariant spectrum.  After dark matter
(DM) begins to dominate the energy density of the Universe, these perturbations
begin to collapse, forming a hierarchical spectrum of DM structures
today.  This spectrum is predicted to extend to subhalo masses well below those of
dwarf Galaxies, $M_\text{sh} \sim 10^9\,M_\odot$, which are the least-massive
DM subhalos observed so far.  Discovering DM subhalos with even lower mass is
complicated by the fact that such objects are not expected to host many stars.
Finding such subhalos is of the utmost importance for a number of reasons: {\it (i)}
their existence is a so-far untested prediction of
$\Lambda$CDM~\cite{Springel:2008cc}, {\it (ii)} certain particle and cosmological
models for DM, including warm DM~\cite{Colin:2000dn,Bode:2000gq,Viel:2005qj},
fuzzy DM~\cite{Hu:2000ke,Li:2013nal,Hui:2016ltb}, and self-interacting
DM~\cite{Spergel:1999mh,Wandelt:2000ad,Kaplinghat:2013kqa},
predict drastic
deviations from the $\Lambda$CDM prediction for the halo-mass function, which
describes the number of halos as a function of mass, at scales below the dwarf
scale, and {\it (iii)} low-mass and nearby subhalos could provide invaluable targets
for indirect searches for DM annihilation and decay.  

In this work, we propose a novel method for finding low-mass DM subhalos.
DM subhalos perturb the phase-space distribution of stars as they propagate
through the local Galaxy.  These perturbations, which we dub ``stellar wakes",
are a key signature of low-mass subhalos that may be observable with upcoming
data from {\it e.g.}\ the {\it Gaia} mission~\cite{2001A&A...369..339P}, the
Large Synoptic Survey Telescope (LSST)~\cite{2009arXiv0912.0201L}, and the
Dark Energy Survey (DES)~\cite{2005astro.ph.10346T}, combined with existing
surveys from, for example, the Sloan Digital Sky Survey
(SDSS)~\cite{Juric:2005zr}.  In the left panel of Fig.~\ref{fig:phase}, we
show an example of the perturbed stellar phase-space distribution caused by a
$\sim 2 \times 10^7\,M_\odot$ subhalo.  Stars are pulled towards the subhalo
as it passes, leaving behind distinct features in their velocity distribution
and, to a lesser extent, in their number density distribution.
 
\begin{figure*}[htb]
  \includegraphics[width=\textwidth]{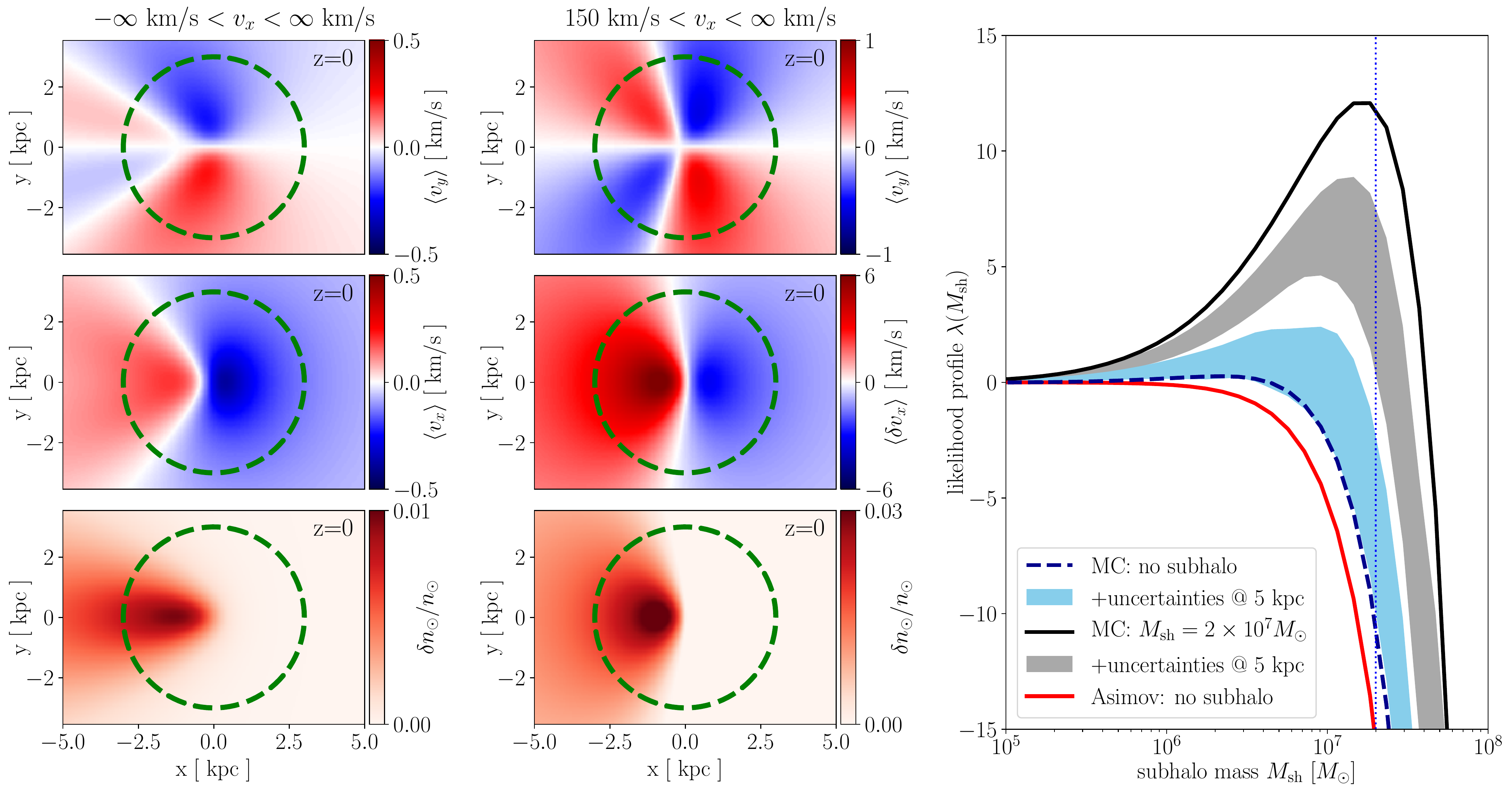}
  \vspace{-.50cm}
  \caption{{\it (Left)} Moments of the stellar phase-space distribution (at $z
    = 0$ kpc), perturbed by a passing DM subhalo at the origin. The subhalo is
    described by a Plummer sphere with
    $M_\text{sh} = 2 \times 10^{7}\,M_\odot$, $r_s \approx 0.72~\text{kpc}$,
    and is traveling in the $\vec{\hat{x}}$ direction with
    $v_\text{sh} = 200~\text{km/s}$.
    The background phase-space distribution is
    described by a Maxwell--Boltzmann distribution (see~\eqref{f0}) with $v_0 =
    100~\text{km/s}$.
    {\it (Middle)} Same as left, but selecting only stars
    that are co-moving with the subhalo with $v_x>150$~km/s.
    {\it (Right)}
    The stellar-wakes likelihood profile, defined in~\eqref{LL_phase}, as a
    function of the assumed subhalo mass. We show results for a simulation with
    the same background and subhalo parameters as in the left panel (black solid)
    and for a background-only simulation without a subhalo (blue dashed). The
    corresponding uncertainties, described in the text, are shown in
    gray/light blue, respectively, for a distance of 5~kpc
    from the center of the ROI to Earth.  The
    unperturbed stellar number density is $n_0 = 5\times10^3$~kpc$^{-3}$, and we
    use an ROI with radius $R = 3$~kpc (dashed green in the left panel). For the
    simulation including a subhalo, the likelihood profile peaks at the correct
    subhalo mass, marked by a vertical dotted blue line.  For the background-only
    simulation, we find good agreement with analytic results based on the Asimov
    data set (red), see~\eqref{Asimov_Plummer_limit}.
  }
  \vspace{-0.15in}
  \label{fig:phase}
\end{figure*}

The method proposed here complements the two main techniques in the literature
for searching for low-mass subhalos (see also~\cite{Feldmann:2013hqa}): the
stellar stream method and strong gravitational lensing.  As subhalos pass by
cold stellar streams, they perturb the phase-space distribution of stars in the
streams, and these perturbations may expand into relatively large gaps. The
stellar stream method may be able to probe the halo-mass function at subhalo
masses down to $M_\text{sh} \sim 10^5 \, M_\odot$~\cite{2002MNRAS.332..915I,
2002ApJ...570..656J, SiegalGaskins:2007kq, Bovy:2015mda,2016ApJ...820...45C,
2016MNRAS.463..102E, 2017MNRAS.470...60E}.  In fact, two gaps recently
identified in the Pal~5 stellar stream~\cite{2017MNRAS.470...60E} may originate
from $\sim 10^6$--$10^8\,M_\odot$ subhalos.  However, it is hard to
conclusively interpret gaps in stellar streams as arising from DM subhalos,
since the DM subhalos are no longer expected to be present near the stream.
For example, the two gaps in~\cite{2017MNRAS.470...60E} could also have arisen
from perturbations due to the Mily Way's bar or passing giant molecular clouds.
Another method to detect subhalos is using strong lensing of distant
galaxies~\cite{Mao:1997ek} (see also~\cite{2013ApJ...767....9H} and references
therein).  Recent strong lensing observations using the Atacama Large
Millimeter/submillimeter Array~\cite{Hezaveh:2016ltk} have already claimed
detection of DM subhalos at masses $\sim 10^8$--$10^9 \, M_\odot$.

The advantage of the stellar wakes method proposed here is that it could detect
nearby DM subhalos, potentially at masses down to $\sim 10^7$ $M_\odot$ using
halo stars and masses $\sim 10^6 \, M_\odot$ using disk stars. Moreover, it
pinpoints the current subhalo positions, enabling detailed followup studies and
even searches for DM annihilation and decay.

In the following, we will calculate the modification to the stellar phase-space
distribution from passing subhalos analytically and then develop a likelihood
framework to search for DM subhalos.  We will validate this framework on
simulated stellar populations, including projected observational uncertainties,
and we will discuss potential applications.

%---------------------------------------------------------------------------

\vspace{1ex}
{\bf Perturbed stellar phase-space.}---We assume that a local stellar
population is in kinetic equilibrium such that, within the region of interest
(ROI) where we will perform the analysis, its phase-space distribution
may be described by a homogeneous, time-independent
distribution $f_0(\vec{v})$, with $\vec{v}$ the stellar velocities.  The
number density is given by \mbox{$n_0 = \int \! d^3v \, f_0(\vec{v})$}.  The
gravitational potential of a passing subhalo induces an out-of-equilibrium
perturbation to the phase-space distribution, which we write as
\es{fexp}{
  f(\vec{x}, \vec{v}, t) = f_0(\vec{v}) + f_1(\vec{x}, \vec{v}, t) \,.
}
In general, the phase-space distribution is a solution to the collisionless
Boltzmann equation
\es{boltz_0}{
  {\partial f \over \partial t}  +  \vec{v} \cdot \vec{\nabla}_x f
                                 -  \vec{\nabla}_x \Phi \cdot \vec{\nabla}_v f = 0 \,,
}
where $\Phi$ is the gravitational potential generated by the subhalo.
By substituting~\eqref{fexp} into~\eqref{boltz_0}, we may derive
the equations of motion for $f_1$.  We choose to do so perturbatively,
expanding to leading order in Newton's constant $G$. In this approximation,
the term $\nabla_x \Phi \cdot \nabla_v f_1$, which would be of order
$G^2$, can be dropped~\cite{Brandenberger:1987kf}:
\es{boltz_1}{
  {\partial f_1 \over \partial t} + \vec{v} \cdot \vec{\nabla}_x f_1
    = \vec{\nabla}_x \Phi \cdot \vec{\nabla}_v f_0  \,.
}
We will
work, for now, in the subhalo rest frame, where $\Phi$ is time-independent
and thus the velocity distribution $f_1$ is static. 
In~\cite{Brandenberger:1987kf} it was shown, by first going to
Fourier space for the variable $\vec{x}$, that the solution to~\eqref{boltz_1}
is given by 
\es{BKT_gen}{
  f_1(\vec{x}, \vec{v}) = \int_0^\infty {du \over u^2}
                          \vec{\nabla}_y \Phi(\vec{y}) \cdot \vec{\nabla}_{v} f_0(\vec{v})
                          \bigg|_{\vec{y} = \vec{x} - \vec{v}/u} \,.
}
Throughout this work, we take $f_0(\vec{v})$ to be a boosted Maxwell-Boltzmann
distribution of the form
\es{f0}{
  f_0( \vec{v} ) = {n_0 \over \pi^{3/2} v_0^3}
                    e^{-(\vec{v} + \vec{v}_\text{sh} )^2 / v_0^2} \,,
}
in the subhalo rest frame.
Here, $v_0$ is the velocity dispersion and $\vec{v}_\text{sh}$ is the boost of
the subhalo with respect to the frame where the velocity distribution is
isotropic (for example, the Galactic frame).  The generalization to velocity
distributions with anisotropic velocity dispersions is straightforward, but the
isotropic case suffices for illustration.  We model the density profiles of DM
subhalos within the inner regions of the Milky Way (MW) by ``Plummer spheres''
\cite{Plummer:1911zza} with
\es{eq:rho-plummer}{
  \rho(r) = \frac{3 M_\text{sh}}{4\pi r_s^3} \Big(1 + \frac{r^2}{r_s^2} \Big)^{-5/2} \,,
  \quad
  \Phi(r) = - \frac{G M_\text{sh}}{\sqrt{r^2 + r_s^2}} \,.
}
The Plummer profile features a constant density core of characteristic radius
$r_s$.  At large radii, it drops off faster than the standard
Navarro--Frenk--White (NFW) profile~\cite{Navarro:1996gj}, reflecting the tidal
stripping which is expected to occur in the outer layers of field halos within
the MW. The Plummer model also has the advantage of being easier to work
with analytically than the NFW profile.  We use the results
of~\cite{2016MNRAS.463..102E}, which analyzed subhalos within the Via Lactea II
simulation~\cite{Diemand:2008in}, to estimate the mass dependence of $r_s$:
$ r_s \approx 1.62~\text{kpc} \times (M_\text{sh} / 10^8 M_\odot)^{1/2}$.
From~\eqref{BKT_gen}, we then obtain
\es{eq:f1-plummer}{
  f_1(\vec{x}, \vec{v})
    = &\frac{2 G M_\text{sh} n_0}{\pi^{3/2} v_0^5} e^{-(\vec{v} + \vec{v}_\text{sh})^2 / v_0^2}
      (\vec{v} + \vec{v}_\text{sh}) \\
      &\times \frac{\sqrt{1 + \frac{r_s^2}{x^2}} \vec{\hat{v}} - \vec{\hat{x}}}
           {v \, x \, \sqrt{1 + \frac{r_s^2}{x^2}} \,
             \big( \sqrt{1 + \frac{r_s^2}{x^2}} - \vec{\hat{x}} \cdot \vec{\hat{v}} \big)} \,,
}
for a subhalo located at the origin.
 
In the left column of Fig.~\ref{fig:phase} we illustrate three different
analytically determined moments of the stellar phase-space distribution,
perturbed according to~\eqref{eq:f1-plummer} by a subhalo with the characteristics
given in the caption.  In particular, we show the average velocities $\ev{v_x}$
and $\ev{v_y}$ in the $\vec{\hat x}$ and $\vec{\hat y}$
directions, respectively, as well as the fractional change in the number
density $\delta n_\odot / n_\odot$.  All three panels show slices of
phase-space in the $x$--$y$ plane at $z = 0~\text{kpc}$.  
To illustrate that perturbations are largest for comoving stars, we show in the
center column similar distributions using only stars that satisfy
$v_x>150$~km/s.  This cut selects stars that are moving with the subhalo and
are therefore perturbed the most by its presence.  Since the selection
$v_x>150$~km/s implies non-zero $\ev{v_x}$, we show $\ev{\delta v_x} \equiv
\ev{v_x - \text{175~km/s}}$ instead of $\ev{v_x}$.

%---------------------------------------------------------------------------

\vspace{1ex}
{\bf Stellar wakes likelihood function.}---Given kinematic data on a large
population of stars, we can use~\eqref{eq:f1-plummer} to search for
evidence of a DM subhalo.  We stress that we are not looking for stars bound to
the DM subhalo but rather for a perturbation to the ambient
distribution of stars consistent with the expectation from a passing
gravitational potential.  

We will focus here on a formalism that utilizes the full 6-D kinematic data for
the stellar population.  This requires a complete sample of stars in order to
not introduce bias in the spatial dependence of the number density.  In the
Supplementary Material we show that similar results are obtained using a
likelihood function based only on the velocity distribution; this may be the
preferred method if a complete sample of stars is not available, as long as an
unbiased determination of the velocity distribution is possible.

The un-binned likelihood function is given by 
\es{LL_phase}{
  p(d | {\mathcal M}, {\bm \theta}) &= e^{- N_\text{star}({\bm \theta})}
    \prod_{k=1}^{\bar N_\text{star}} \,  f(\vec{x}_k, \vec{v}_k)({\bm \theta}) \,,
}
where
\es{Numbe}{
  N_\text{star}({\bm {\theta}}) \equiv \int_\text{ROI} \! d^3\vec{x} \, d^3\vec{v}
                                       f(\vec{x},\vec{v})({\bm \theta}) 
} 
is the total predicted number of stars in the ROI, as a function of the
model parameters ${\bm \theta}$ in the model $\mathcal{M}$.  The product
is over all $\bar N_\text{star}$ stars within the ROI; their kinematic
parameters $\{ \vec{x}_k, \vec{v}_k \}$ form the data set $d$.
For a spherical ROI of radius $R$ and for the Plummer model,
$N_\text{star}({\bm {\theta}})$ is given by
\es{eq:f1-norm-plummer-2}{
  N_\text{star}({\bm {\theta}}) = &{4 \over 3} \pi R^3 n_0
    \bigg[ 1 +  3 G M_\text{sh} \; F\Big( {v_\text{sh} \over v_0} \Big) \\[0.2cm]
   &\times \frac{R \sqrt{R^2 + r_s^2} - r_s^2 \sinh^{-1} (R/r_s)}
                     {R^3 \, v_0 v_\text{sh}} \bigg]\,,
}
where $F(x)$ is the Dawson integral.  The model parameters ${\bm\theta}$ include the
parameters $n_0$ and $v_0$ of the background distribution $f_0$, in addition to
the DM subhalo parameters $M_\text{sh}$, $r_s$, its position
$\vec{x}_\text{sh}$, and its boost $\vec{v}_\text{sh}$.

With large numbers of stars, it may be easier to use the binned likelihood
\es{LL_phase_binned}{
  p(d | {\mathcal M}, {\bm \theta}) &= 
    \prod_{i=1}^{N_\text{bins}} \, e^{-n_i(\bm\theta)}
    \frac{[n_i(\bm\theta)]^{N_i}}{N_i!} \,.
}
Here, the observed number of stars in each of
the $N_\text{bins}$ 6-D phase space bins is denoted by $N_i$. The corresponding model
prediction is given by \mbox{$n_i({\bm\theta}) \equiv d^3\vec{x} \, d^3\vec{v} \,
f(\vec{x}_i, \vec{v}_i)({\bm\theta})$},
where $d^3\vec{x} \, d^3\vec{v}$ is the phase space volume covered by each bin,
and $\vec{x}_i$, $\vec{v}_i$ are its 6-D coordinates.

To set a constraint on
$M_\text{sh}$, fixing or marginalizing over the other subhalo parameters and
background parameters, we construct the likelihood profile
\es{LL}{
  \lambda(M_\text{sh}) = 2
    \big[ &\max_{{\bm\theta}_\text{nuis}} \log p(d | {\mathcal{M}}, {\bm\theta} )
   \\
    &- \max_{\bm\theta}  \log p(d | {\mathcal{M}}, {\bm\theta}) \big] \,.
}
Here, we have written ${\bm\theta} =
\{M_\text{sh}, {\bm\theta}_\text{nuis} \}$, where ${\bm\theta}_\text{nuis}$
denotes the rest of the subhalo and background nuisance parameters.  The 95\%
upper bound on $M_\text{sh}$ is 
determined by $\lambda(M_\text{sh}^{95}) \approx -2.71$, with $M_\text{sh}^{95}$ greater than the mass that maximizes the likelihood~\cite{Cowan:2010js}.
We may use the same framework to estimate the significance of a
detection in the event that a subhalo is present in the data.  In this case, it
is useful to define a test statistic (TS) given by twice the maximum log-likelihood
difference between the models with and without the DM subhalo:
\es{TS}{
  \text{TS} = 2 \big[ &\max_{\bm\theta} \log p(d | {\mathcal{M}}, {\bm\theta})
     \\
    &-\max_{{\bm\theta}_\text{nuis}}  \log p(d | {\mathcal{M}}, {\bm\theta})
      \big|_{M_\text{sh} = 0} \big] \,.
}
  
To estimate the sensitivity of the stellar wakes likelihood to the presence of
DM subhalos, we use the Asimov data set~\cite{Cowan:2010js}, which corresponds to
the median stellar phase space distribution that would be
obtained over many realizations of mock data.  For the binned likelihood from
Eq.~\eqref{LL_phase_binned}, it is given by
$N_i = d^3\vec{x} \, d^3\vec{v} \, f_0(\vec{v}_i)$. 
The Asimov framework allows us to analytically estimate the median likelihood profile that would be obtained over multiple Monte Carlo simulations.
Expanding to leading order in Newton's constant
we find 
\begin{align}
  \lambda(M_\text{sh}) &\approx
    -\int \! d^3\vec{x} \, d^3\vec{v} {f_1^2(\vec{x}, \vec{v}) \over f_0(v)} \,.
  &\text{(Asimov)}
  \label{Asimov_Limit}
\end{align}
For the Plummer sphere model, and a spherical ROI of radius R, we may calculate 
\begin{align}
  \lambda(M_\text{sh}) &\approx
    -{64 \pi \,n_\odot G^2 M_\text{sh}^2 R \over v_0^2 v_\text{sh}^2}
    {\mathcal I}(\epsilon_v, \epsilon_r) \,,
  &\text{(Asimov)}
  \label{Asimov_Plummer_limit}
\end{align}
where $\epsilon_v = v_0 / v_\text{sh}$ and $\epsilon_r = r_s / R$. An integral
expression for ${\mathcal I}(\epsilon_v, \epsilon_r)$ is given in the
Supplementary Material.  ${\mathcal
I}(\epsilon_v, \epsilon_r)$ is close to unity at $\epsilon_v, \epsilon_r \ll 1$
and falls quickly for $\epsilon_v \gtrsim 1$ and $\epsilon_r \gtrsim
0.5$.   
For the Asimov dataset, the test statistic is given by $\text{TS} =
-\lambda(M_\text{sh})$, so that~\eqref{Asimov_Limit}
and~\eqref{Asimov_Plummer_limit} may be used to estimate the sensitivity to
detection as well as exclusion ($5\sigma$ detection corresponds
to $\text{TS} \approx 25$).

%---------------------------------------------------------------------------

\vspace{1ex}
{\bf Simulation results.}---It is useful to verify the above formalism on
simulated data.  We generate a homogeneous population of halo
stars from a phase space distribution with $v_0 = 100~\text{km/s}$ and $n_0 =
5\times10^3 / \text{kpc}^3$, consistent with the number density of blue stars
measured by SDSS~\cite{Juric:2005zr} far away from the disk at $\sim 8$~kpc from
the Galactic Center.
We then simulate the stellar trajectories
in the presence of a DM subhalo described by a Plummer sphere with
$M_\text{sh} = 2 \times 10^{7}\,M_\odot$, $r_s \approx 0.72~\text{kpc}$,
and traveling in the $\vec{\hat{x}}$ direction with $v_\text{sh} = 200~\text{km/s}$.
The subhalo is initially far away from
the spherical ROI with radius $R = 3~\text{kpc}$, and we end the simulation
when it reaches the center of the ROI at $(x,y,z) = (0,0,0)$~kpc.
Note that, while simulating
the stellar trajectories in the gravitational potential of the subhalo, we
ignore the potential generated by the stars themselves.  

In the right panel of Fig.~\ref{fig:phase}, we show the 1-D likelihood profile
as a function of $M_\text{sh}$ for the likelihood analysis
performed on the simulated data.
The TS defined in~\eqref{TS} (black line) favors the presence of a subhalo with
the correct mass (dotted blue line) over the
background-only hypothesis at a value $\text{TS} \approx 12$.
This matches the expectation based on the likelihood profile from the
Asimov analysis (Eq.~\eqref{Asimov_Plummer_limit}, shown in red), which in turn
agrees with the likelihood profile constructed on a control
simulation sample without a subhalo (dashed blue).

Observational uncertainties on stellar kinematic data can alter the likelihood
profiles, as they tend to artificially increase, for example, the velocity
dispersion and smear localized structure. 
This is illustrated in the right panel of
Fig.~\ref{fig:phase}, using a proposed observational setup similar to that
taken in~\cite{2015MNRAS.454.3542E}.  
We assume that the sky position uncertainties are similar to those
projected for {\it Gaia}~\cite{2010A&A...523A..48J}
and at the level of a few $\muas$ for bright stars and a few hundred $\muas$ at the dim 
end~\cite{2010A&A...523A..48J}.
At distances of a few kpc from Earth, these uncertainties are expected to be subdominant compared to the distance uncertainties, determined by photometric parallax.
DES, for example, uses the photometric parallax method with very small $r$-band magnitude uncertainties, though there is still an intrinsic photometric scatter $\sim 0.3$ mag; this translates into a distance uncertainty $\sim 14$\%.

The proper motion can be measured accurately by {\it Gaia} with uncertainties
similar to the position on the sky, but the conversion to physical velocity
involves the distance of the star and thus that source of uncertainty also
plays an important role.  For radial velocities, we assume an uncertainty of
5\,km/s, although we expect many stars to be measured more
accurately~\cite{2015MNRAS.454.3542E} by surveys such as
VLT~\cite{2011ApJ...736..146K,2012Msngr.147...25G},
WEAVE~\cite{doi:10.1117/12.925950}, and 4MOST~\cite{doi:10.1117/12.2232832}.
In our simulations, velocity uncertainties play a sub-dominant role compared to
position uncertainties.

We include 68\% confidence bands in Fig.~\ref{fig:phase}, where we marginalized
over different Monte Carlo realizations and subhalo velocity directions with
respect to the line of sight, assuming a distance of 5 kpc from Earth with
uncertainties mentioned above.  The significance for a subhalo is slightly
reduced and the TS is artificially enhanced when there is no subhalo present
due to the observational uncertainties.  Note that the impact of statistical
jitter on parallax measurements is highly asymmetric to the extent that the
unperturbed curves in Fig.~\ref{fig:phase} are not contained in the 68\% containment
bands.

We remark that future surveys such as LSST could significantly increase the
number density of stars available for such an analysis, by allowing for dimmer and
redder stars, which would lead to an enhanced sensitivity to lower-mass
subhalos in the stellar halo.

%---------------------------------------------------------------------------

\vspace{1ex}
{\bf Discussion.}---We have presented a novel method for identifying low-mass
DM subhalos, potentially down to $\lesssim 10^7 \, M_\odot$, through their
effects on halo stars.  The method requires a large sample of stellar kinematic
data, which may be available with upcoming surveys such as those by {\it Gaia}
and LSST.  

To estimate the chances of finding a suitable subhalo target for such
observations within our local Galactic neighborhood, we estimate the number of
subhalos with $M_\text{sh} \gtrsim 10^7 \, M_\odot$ within a 10~kpc spherical
region around the MW to be $\sim 1.5$.  This estimate arises from assuming a
local halo-mass function \mbox{$dN/(dM_\text{sh} dV) \approx 630~\text{kpc}^{-3}
\, M_\odot^{-1} \times (M_\text{sh} /
M_\odot)^{-1.9}$}~\cite{Hooper:2016cld}, based on an analysis of subhalos in
the ELVIS simulation~\cite{Garrison-Kimmel:2013eoa} and assuming the subhalo
density follows an Einasto distribution~\cite{Springel:2008cc,Despali:2016meh}.
These numbers were derived from DM-only $N$-body simulations; it has been
claimed~\cite{Sawala:2016tlo,2017MNRAS.465L..59E}, using semi-analytic methods
and hydrodynamic simulations, that baryonic effects including increased tidal
forces and disk crossings could reduce the number of subhalos by a factor $\sim
2$.

The detection of $\sim 10^6 \, M_\odot$ subhalos would likely require colder
and denser stellar populations than available in the stellar halo, such as
populations of MW disk stars.  Searches in the disk may be complicated by
additional sources of out-of-equilibrium dynamics, such as stellar
over-densities, molecular clouds, and density waves.  The prospects of
searching for stellar wakes with disk stars deserve further study.

The analysis proposed in this Letter relies on several assumptions that should
be analyzed in more detail.  First, we have assumed that halo stars are
well virialized, an assumption that could break down in
certain parts of the halo, for example due to the presence of stellar
substructure~\cite{Lisanti:2014dva}.  More detailed Galactic-scale simulations
could help address this potential issue.  Moreover, we have assumed that
the background stellar distribution within the ROI is homogeneous; generalizing
our framework to allow for
space-dependent background distributions should be straightforward
and useful for regions near the disk.  An additional effect that could be
important is the gravitational back-reaction of the over-density induced by the
subhalos.  This may be important for subhalos traversing dense regions, such as
those found near the Galactic plane, and for more compact subhalos that induce
larger over-densities.  More compact subhalos than those predicted in standard
cosmology, such as ultra-compact minihalos, could arise from phase-transitions
in the early Universe or a non-standard spectrum of density perturbations on
small scales generated from dynamics towards the end of
inflation~\cite{Aslanyan:2015hmi}. 

One potential way of testing the stellar wakes formalism could be to utilize
nearby globular clusters, such as 47~Tucanae and Omega Centauri with masses
$\gtrsim 10^6$ $M_\odot$, as targets.  Globular clusters are more compact than
DM subhalos and are often located near the Galactic plane, where stellar number densities
are higher than in the halo.

An open-source code package for performing the likelihood analysis presented in
this Letter, along with example simulated datasets, is available at
\url{https://github.com/bsafdi/stellarWakes}.

\noindent
{\bf Acknowledgments.}---We thank Vasily Belokurov, Tongyan Lin, Mariangela
Lisanti, Lina Necib, Annika Peter, Nick Rodd, Katelin Schutz, and Tracy Slatyer
for useful discussions.  BRS especially thanks Tongyan Lin and Marilena LoVerde
for collaboration on unpublished related work.  The work of MB and JK
has been supported by the German Research Foundation (DFG) under Grant Nos.\
EXC-1098, \mbox{KO~4820/1--1}, FOR~2239 and GRK~1581, and by the European
Research Council (ERC) under the European Union's Horizon 2020 research and
innovation programme (grant agreement No.\ 637506, ``$\nu$Directions'').  C.-L.
Wu is partially supported by the U.S. Department of Energy under grant Contract
Numbers DE-SC00012567 and DE-SC0013999 and partially supported by the Taiwan
Top University Strategic Alliance (TUSA) Fellowship.  This work was performed in part at the Aspen Center for Physics, which is supported
by NSF grant PHY-1066293.

\twocolumngrid
\def\bibsection{} 
\bibliographystyle{JHEP}
\bibliography{fermi_darksky}

\clearpage
\newpage
\maketitle
\onecolumngrid
\begin{center}
\textbf{\large Stellar Wakes from Dark Matter Subhalos} \\ 
\vspace{0.05in}
{ \it \large Supplementary Material}\\ 
\vspace{0.05in}
{}
{ Malte Buschmann, Joachim Kopp, Benjamin R. Safdi, and Chih-Liang Wu}

\end{center}
%%%%%%%%%% Merge with supplemental materials %%%%%%%%%%
\counterwithin{figure}{section}
\counterwithin{table}{section}
\counterwithin{equation}{section}
\setcounter{equation}{0}
\setcounter{figure}{0}
\setcounter{table}{0}
\setcounter{section}{0}
\renewcommand{\theequation}{S\arabic{equation}}
\renewcommand{\thefigure}{S\arabic{figure}}
\renewcommand{\thetable}{S\arabic{table}}
\newcommand\ptwiddle[1]{\mathord{\mathop{#1}\limits^{\scriptscriptstyle(\sim)}}}

\subsection{Phase-space distribution}
The aim of this section is to derive explicitly the perturbed phase-space distribution $f_1( \vec{x}, \vec{v})$ and the predicted number of stars $N_{star}(\theta)$ in~\eqref{BKT_gen} and~\eqref{Numbe}, respectively. We first focus on the slightly simpler point mass approximation before evaluating the same quantities for the Plummer sphere profile. 

\subsubsection{Point mass approximation}
The point mass approximation implies $\rho(r) = M_\text{sh} \delta^3(r)$ and $\Phi(r) = - {G M_\text{sh} / r}$. Substituting this and \eqref{f0} into \eqref{BKT_gen} yields directly
\es{f1_BKT_PM}{
f_1( \vec{x}, \vec{v})   &=  -{2 G M_\text{sh} \over  \pi^{3/2} v_0^5} {e^{-(\vec{v} + \vec{v}_\text{sh})^2 / v_0^2}} { ( \vec{v} + {\bf v_\text{sh}} )  \cdot \left( {\bf \hat x} - {\bf \hat v} \right) \over v \, x \left(1 - {\bf \hat x} \cdot {\bf \hat v} \right)}  \,.
}
For $N_{star}(\theta)$ it is beneficial, instead of integrating \eqref{f1_BKT_PM} over the ROI, to rewrite \eqref{Numbe} in terms of the density profile $\rho(r)$, 
\es{BKT_gen0}{
N_\text{star}({\bm {\theta}}) &\equiv \int_\text{ROI} \! d^3\vec{x} \, d^3\vec{v}
                                      \left[ f_0(\vec{v})({\bm \theta}) 
                                      + \int_{0}^\infty {du \over u^2} \nabla_y \Phi( {\bf y}) \cdot \nabla_{v} f_0({\bf v}) \right]_{ {\bf y} = {\bf x} -  {\bf v}  / u } \\
&=  \int_\text{ROI} \! d^3\vec{x} \left[ n_0 +\int_{0}^\infty {du \over u^3} \int d^3 {\bf v} \nabla_y^2 \Phi( {\bf y})  f_0({\bf v}) \right]_{ {\bf y} = {\bf x} -  {\bf v}  / u }\\
&=  \int_\text{ROI} \! d^3\vec{x} \left[ n_0 + 4 \pi G \int_{0}^\infty {du \over u^3} \int d^3 {\bf v} \rho( |{\bf x} -  {\bf v}  / u|) f_0({\bf v})  \right]\,.}  
In the last step we used the relation $\nabla^2_y \Phi( {\bf y}) = 4 \pi G \rho(y)$. For a spherical ROI with radius $R$ around the subhalo, together with \eqref{f0}, the equation above yields
\es{BKT_genPoint}{
N_\text{star}({\bm {\theta}}) &=  \int_\text{ROI} \! d^3\vec{x}\, n_0 \left[ 1 + 
{2 G M_\text{sh} \over {x \,  v_0^2}} e^{ - \left[ 1 -  ({ \bf \hat x} \cdot {\bf \hat v_\text{sh}})^2  \right]{v_\text{sh}^2 / v_0^2} }\ \text{erfc} \left( {v_\text{sh} \over v_0 }{ \bf \hat x} \cdot {\bf \hat v_\text{sh} } \right) \right] \\
&\approx \frac{4}{3}\pi R^3 n_0 \left[ 1+\frac{3 G M_{sh}}{R\,  v_0  \, v_\text{sh}} \; F\left({v_\text{sh} \over v_0}\right)\right],
}  
where $F(x)$ is the Dawson integral $F(x) \equiv e^{-x^2} \int_0^x \!
e^{y^2} dy$. The last steps assumes the approximation ${\bf \hat x} \cdot {\bf \hat v_\text{sh}} < - {v_0 / v_\text{sh}} $ such that $\text{erfc}\left( {v_\text{sh} \over v_0 }{ \bf \hat x} \cdot {\bf \hat v_\text{sh} } \right)\approx 1$.

\subsubsection{Plummer sphere profile}
Using the potential and density profile for the Plummer sphere given in \eqref{eq:rho-plummer}, we can directly reproduce \eqref{eq:f1-plummer} by evaluating the integral in \eqref{BKT_gen}.
For $N_\text{star}(\theta)$ we can again use the expression given in \eqref{BKT_gen0}. However, because the density profile is more complex, we need to additionally express $\rho(|\vec{x} - \vec{v}/u|)$ and $f_0(\vec{v})$ in terms of their Fourier transforms
\es{PlummerFourier}{
  \tilde{\rho}(\vec{p}) &\equiv \int\!d^3\vec{x} \, \rho(\vec{x}) e^{i \vec{p}\cdot\vec{x}}
                          = M p \, r_c K_1(p r_c) \,, \\
  \tilde{f_0}(\vec{p})  &\equiv \int\!d^3\vec{v} \, f_0(\vec{v}) e^{i \vec{p}\cdot\vec{v}}
                          =      e^{-\frac{1}{4} p^2 v_0^2 + i \vec{p}\cdot\vec{v}_\text{sh}} \,,
}
where $K_1(x)$ is the modified Bessel function of the second kind. This yields
\es{PlummerFourierNStar}{
N_\text{star}({\bm {\theta}}) &= \int_\text{ROI} \! d^3\vec{x} \left[ n_0 + 4 \pi G \int \! d^3\vec{w}\ d^3\vec{p}\ d^3\vec{k} \ 
   \int \!  \frac{du}{u^3} \tilde{f}_0(\vec p) \tilde{\rho}(\vec k)
    e^{-i u \vec{p}\cdot(\vec{w} - \vec{v}_\text{sh}/u + \vec{x})}
    e^{-i \vec{k}\cdot \vec{w}} \right]  \\
  &= \frac{4}{3}\pi R^3 n_0+
    \frac{2 G}{\pi} \int \! d^3\vec{w}\ d^3\vec{p}\ d^3\vec{k}\int \!  \frac{du}{u^3}
    \ \tilde{f}_0(\vec p) \tilde{\rho}(\vec k)
    \frac{\sin(p R u) - p R u \cos(p R u)}{p^3} e^{-i u \vec{p}\cdot (\vec{w} - \vec{v}_\text{sh}/u)}
    e^{-i \vec{k}\cdot \vec{w}} \\
  &= \frac{4}{3}\pi R^3 n_0+
    \frac{2 G}{\pi} \int \! d^3\vec{p}\int \!  \frac{du}{u^3}
    \tilde{f}_0(\vec p) \tilde{\rho}(u \vec p)
    \frac{\sin(p R u) - p R u \cos(p R u)}{p^3}
    e^{i \vec{p}\cdot \vec{v}_\text{sh}} \,.
}
When we now evaluate first the integral over $u$, then the integral over $\vec{p}$, we can reproduce \eqref{eq:f1-norm-plummer-2}, 
where $F(x)$ is again the Dawson integral $F(x) \equiv e^{-x^2} \int_0^x \!
e^{y^2} dy$.

\subsection{Asimov limit}
\begin{figure}[htb]
  \includegraphics[width=0.5\columnwidth]{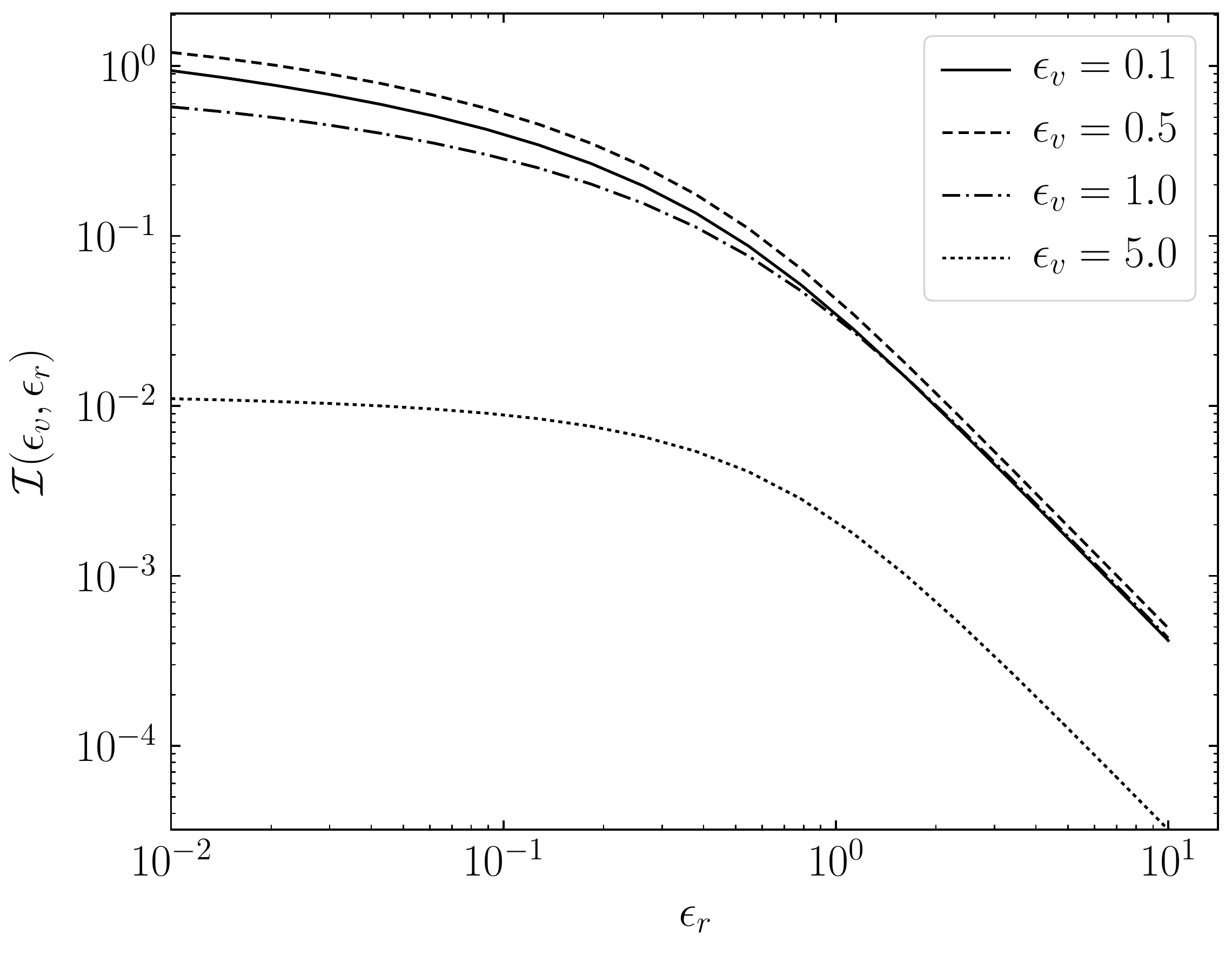}
  \vspace{-.20cm}
  \caption{The function $\mathcal{I}(\epsilon_v, \epsilon_r)$ entering
    the Asimov likelihood profile in~\eqref{Asimov_Plummer_limit} for
    the Plummer sphere model. The parameters are $\epsilon_v = v_0 / v_\text{sh}$ and
    $\epsilon_r = r_s / R$.  Larger values of $\mathcal{I}(\epsilon_v, \epsilon_r)$
    indicate better sensitivity to DM subhalos.
  }
  \vspace{-0.15in}
  \label{fig:I}
\end{figure}

To derive the leading order expression for the likelihood profile in \eqref{Asimov_Limit} it is easiest to start with the binned likelihood in \eqref{LL_phase_binned}. We have to take the logarithm of this quantity,
\es{Log1}{
\log p(d | {\mathcal M}, {\bm \theta}) &\approx \sum_{i=1}^{N_\text{bins}}\left[ -n_i(\theta) + N_i\log n_i(\theta) - N_i(\log N_i-1)\right] \,,
}
which in the continuous limit reads
\es{Log2}{
\log p(d | {\mathcal M}, {\bm \theta}) &= \int \! d^3\vec{x} \, d^3\vec{v}\left[ -f(\vec{x}, \vec{v})(\theta) + f_0(\vec{v})\log f(\vec{x}, \vec{v})(\theta) - f_0(\vec{v})(\log f_0(\vec{v})-1)\right] \\
&\approx \int \! d^3\vec{x} \, d^3\vec{v}\left[ -f(\vec{x}, \vec{v})(\theta) + f_0(\vec{v})\left(
\log f_0(\vec{v})+\frac{f_1(\vec{x}, \vec{v})(\theta)}{f_0(\vec{v})}-\frac{f_1^2(\vec{x}, \vec{v})(\theta)}{2f_0^2(\vec{v})}
\right) - f_0(\vec{v})(\log f_0(\vec{v})-1)\right] \\
&= - \int \! d^3\vec{x} \, d^3\vec{v} \frac{f_1^2(\vec{x}, \vec{v})(\theta)}{2f_0(\vec{v})} \,.
}
In the second line we expanded $\log f(\vec{x}, \vec{v})(\theta)\equiv \log[f_0(\vec{v}) +f_1(\vec{x}, \vec{v})(\theta)]$ for 
$f_1(\vec{x}, \vec{v})(\theta)\ll f_0(\vec{v})$. Substituting \eqref{Log2} into \eqref{LL} reproduces \eqref{Asimov_Limit}.
When we write out \eqref{Asimov_Limit} for a Plummer sphere profile, we obtain \eqref{Asimov_Plummer_limit} with 
\es{TS_asimov_PS}{
{\mathcal I}(\epsilon_v,\epsilon_r) &=  \frac{1}{\sqrt{\pi}} \int_{-1}^{1} {d(\cos \theta_x) \over 2} \int_0^\infty d \tilde v \, \int_{-1}^{1} {d (\cos \theta_v) \over 2} \int_0^{2 \pi} {d \phi_v \over 2 \pi} \int_{0}^{1} {d\tilde r}
{\tilde v^4 e^{- \tilde v^2}  \over 1 + \epsilon_v^2 \tilde v^2 - 2 \epsilon_v \tilde v \cos \theta_v} \left[ {{\bf \hat v} \cdot \left(   a \vec{\hat{v'}}  - \vec{\hat{x}}    \right) \over a\left(a - {\bf \hat x} \cdot {\bf \hat v'} \right)}  \right]^{2} \,,
}
where $a\equiv\sqrt{1 + \epsilon_r^2/\tilde r^2}$, $\tilde r\equiv r/R$, $\tilde v\equiv v/v_0$, and $\hat{\vec{v}}'\equiv (\vec{v}-\vec{v}_\text{sh})/\sqrt{v^2+v^2_\text{sh}-2\vec{v}\cdot\vec{v}_\text{sh}}$. Note that the integral over $\tilde r$ can be performed analytically, but due to the length of the expression we refrain from quoting it explicitly. All other integral have to be evaluated numerically. In Fig.~\ref{fig:I} we illustrate this function
over a range of relevant $\{\epsilon_v, \epsilon_r\}$.  

\subsection{6-D vs. 3-D kinematic data}

In order to asses the advantage of using the full 6-D kinematic data, as opposed to 3-D information, we need to know the un-binned likelihood functions based on only the velocity or number density. They can be written analogously to \eqref{LL_phase}:
\es{LL_phase_velo}{
  p_\text{velocity}(d | {\mathcal M}, {\bm \theta}) &= 
    \prod_{k=1}^{\bar N_\text{star}} \,  \frac{f(\vec{x}_k, \vec{v}_k)({\vec{\theta}})}{n(\vec{x}_k)(\vec{\theta})} \,,
}
and
\es{LL_phase_number}{
  p_\text{number}(d | {\mathcal M}, {\bm \theta}) &= e^{- N_\text{star}({\vec{\theta}})}
    \prod_{k=1}^{\bar N_\text{star}} \,  n(\vec{x}_k)({\vec{\theta}}) \,,
}
where the data set $d$ is now restricted to $\{\vec{v}_k\}$ and $\{\vec{x}_k\}$, respectively. 

\begin{figure*}[htb]
 	 \includegraphics[width=0.5\columnwidth]{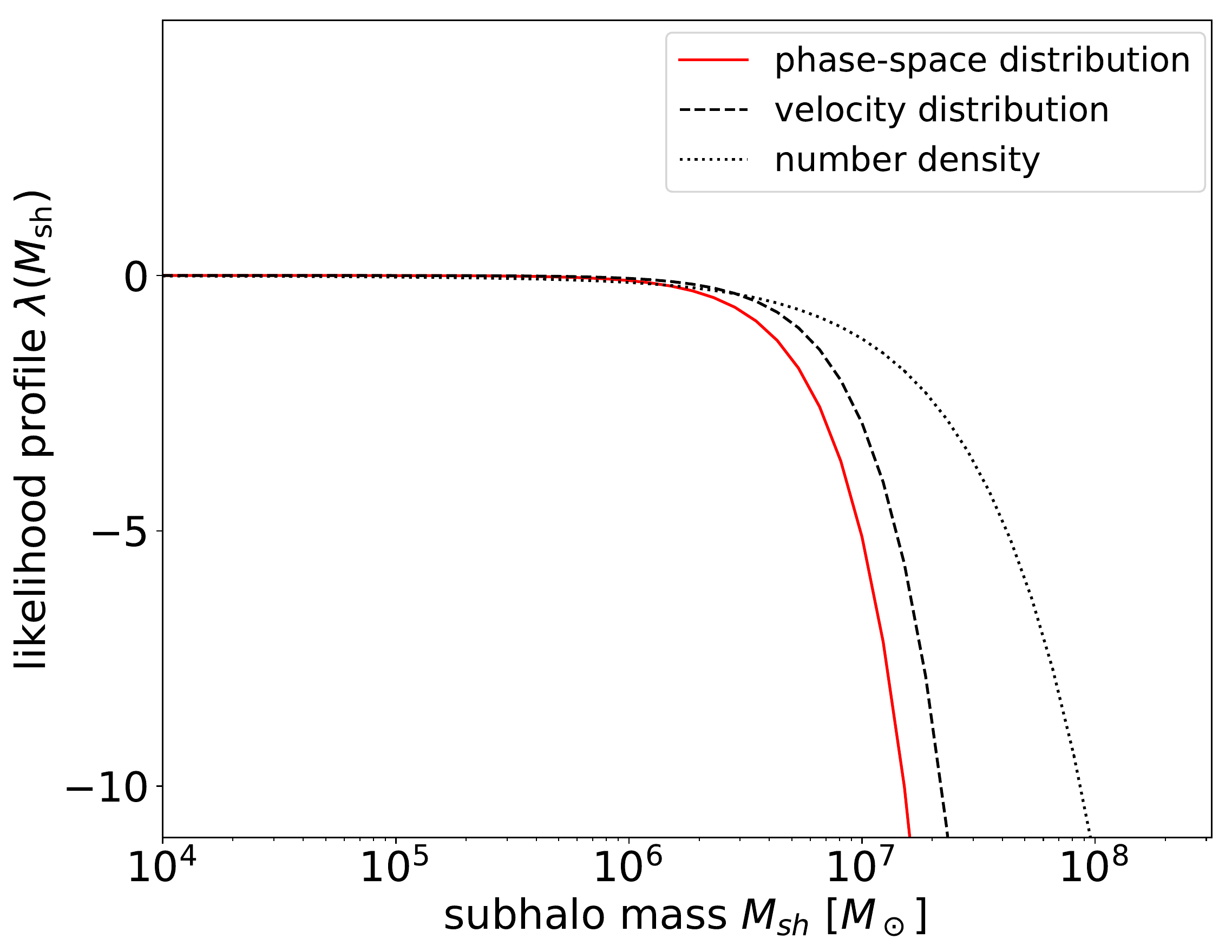}
        \vspace{-.20cm}
        \caption{  Likelihood profiles obtained using the Asimov dataset under the null hypothesis, with parameters as in Fig.~\ref{fig:phase}, for the 6-D and 3-D likelihood functions.  }
        \vspace{-0.15in}
        \label{Fig: TScompare}
\end{figure*}

We show an example of the Asimov likelihood profile, under the null hypothesis, comparing the 6-D and 3-D distributions in Fig.~\ref{Fig: TScompare}. We take $v_0 \,= \,100~  \text{km/s}$, $n_0 \,= \,5\times10^3 \, \text{kpc}^{-3}$, $M_{sh} \,= \,2\times10^7 \, \text{M}_\odot$, $v_\text{sh} \,= \,200 ~ \text{km/s}$, and ROI radius $R \, =\, 3 ~ \text{kpc}$. 
As shown in Fig~\ref{Fig: TScompare}, the likelihood profile using the full phase space information is not much different from that obtained with only velocity information. This indicates that one can work with the simplified likelihood function, which does not require a complete sample of stars, and obtain similar sensitivity to DM subhalos.  On the other hand, the likelihood function that only uses the stellar number density data is significantly less sensitive, by almost one orders of magnitude in mass, to DM subhalos.

\end{document}